\documentclass[aps,prx,reprint,twocolumn,superscriptaddress,english]{revtex4}
\usepackage{amssymb}
\usepackage{graphicx}
\usepackage{amsmath}
\usepackage{amsfonts}
\usepackage[T1]{fontenc}
\usepackage[latin9]{inputenc}
\usepackage{array}
\usepackage{multirow}
\usepackage{color}
\usepackage{esint}
\usepackage{bm}
\usepackage{color}
\usepackage{bbm}
\usepackage{hyperref}
\usepackage{babel}
\usepackage{titlesec}
\usepackage{mathtools}
\DeclarePairedDelimiter\ket{\lvert}{\rangle}

\begin{document}

\title{Collective dipole oscillations of a spin-orbit coupled Fermi gas}

\author{Shanchao Zhang, Chengdong He, Elnur Hajiyev, Zejian Ren, Bo Song and Gyu-Boong Jo$^{*}$}
\affiliation{Department of Physics, The Hong Kong University of Science and Technology,\\ Clear Water Bay, Kowloon, Hong Kong SAR}

\begin{abstract}
The collective dipole mode is induced and measured in a spin-orbit (SO) coupled degenerate Fermi gas of  $^{173}$Yb atoms. Using a differential optical Stark shift, we split the degeneracy of three hyperfine states in the ground manifold, and independently couple consecutive spin states with the equal Raman transitions. A relatively long-lived spin-orbit-coupled Fermi gas, readily being realized with a narrow optical transition, allows to explore a single-minimum dispersion where three minima of spin-1 system merge into and to monitor collective dipole modes of fermions in the strong coupling regime. The measured oscillation frequency of the dipole mode is compared with the semi-classical calculation in the single-particle regime.  Our work should pave the way towards the characterization of spin-orbit-coupled fermions with large spin $s>\frac{1}{2}$ in the strong coupling regime.
\end{abstract}

\maketitle

Spin-orbit (SO) coupling, an essential ingredient for the realization of exotic topological phases, has been extensively investigated in an ultracold atomic system~\cite{Zhai:2015hg,Dalibard:2011gg,Goldman:2014bv} due to its capability of controlling an full atomic Hamiltonian on demand. In an atomic system with a neutral charge, SO coupling is often synthethized by coupling pseudo-spin states via the Raman transition. Pseudo-spins can be chosen in various contexts including internal hyperfine levels, motional states in an optical lattice or the Bloch bands~\cite{Zhang:2018hq}. Nevertheless, most of experimental realization of SO coupling in cold atoms has employed only two spin states owing to the experimental complexity~\cite{Lin:2011hn,Zhang:2012fd,Qu:2013bf,Olson:2014dz,LeBlanc:2013cp,Wang:2012gv,Williams:2013kx,Xu:2013db,Cheuk:2012id,Burdick:2016jt,Song:2016ep,Takasu:2017gq}. Especially, the experimental study of spin-orbit-coupled fermions has been so far limited within the spin-$\frac{1}{2}$ manifold in both bulk and lattice systems in contrast to bosons\cite{Campbell:2016jq,Luo:2016iw}. To realize spin-momentum locking with large spin s$>$$\frac{1}{2}$, multiple Raman transitions need to be set up, which in particular causes severe light-induced heating for alkali fermions.

The recent implementation of SO coupling in non-alkali fermions~\cite{Burdick:2016jt,Song:2016ep,Takasu:2017gq} allows for flexible configuration with SO coupling owing to a narrow optical transition with significantly reduced light-induced heating. So far various schemes have been demonstrated to implement spin-$\frac{1}{2}$ SO coupling with non-alkali fermions in a homogeneous gas~\cite{Song:2016ep}, in an optical Raman lattice~\cite{Song:2017uf,Song:2018vz} and in an optical lattice clock~\cite{Livi:2016cn,Kolkowitz:2017iv}. Here, we extend our capability to engage three hyperfine states that are consecutively coupled through Raman transitions~\cite{Lan:2014fka}. A narrow optical transition of $^{173}$Yb atoms is used to induce Raman coupling with minimal heating. Consequently, we monitor a collective dipole mode in the strong coupling regime where multiple minima merge into the single minimum with long-lived spin-orbit-coupled fermions.

The effect of SO coupling in atomic gases has been extensively investigated. The phase digram is experimentally measured to uncover the (non-)magnetic and stripe phases~\cite{Ji:2014jh}, followed by a recent observation of a stripe superfluid phase in a superlattice with SO coupling~\cite{Li:2016tp}. Furthermore, the dynamic properties such as the collective mode~\cite{Zhang:2012fd}, the excitation spectrum~\cite{Ji:2014jh}, the Zitterbewegung~\cite{LeBlanc:2013cp} and the negative mass~\cite{Khamehchi:2017ku} have been studied with spin-orbit-coupled bosons. Nevertheless, the property of the spin-orbit-coupled fermions still remains to be explored in experiments. Here, we, for the first time, investigate the collective mode of the spin-orbit-coupled degenerate fermions. Although the collective mode measurement has been widely used to understand the properties of atomic gases in various contexts including spin-orbit-coupled bosons~\cite{Zhang:2012fd}, its application to spin-orbit-coupled fermions has remained a challenge due to the limited lifetime of fermions in the presence of SO coupling. In this work, we implement relatively long-lived spin-orbit-coupled fermions , and experimentally measure a collective dipole mode of a spin-orbit-coupled Fermi gas in the collisionless regime. We observe the minimum dipole oscillation frequency around the two-photon detuning $\delta\sim-$2.7$E_r$ at which three spin states are symmetrically coupled in good agreement with the semi-classical theory.  Our observation of the collective dipole mode will pave the way towards not only the spin-orbit-coupled fermions in hydrodynamic regime~\cite{Hou:2015ky} but also quantum dynamics with large spin~\cite{Lan:2014fka}.

\begin{figure}[ptb]
\begin{center}\includegraphics[width=8.5cm]{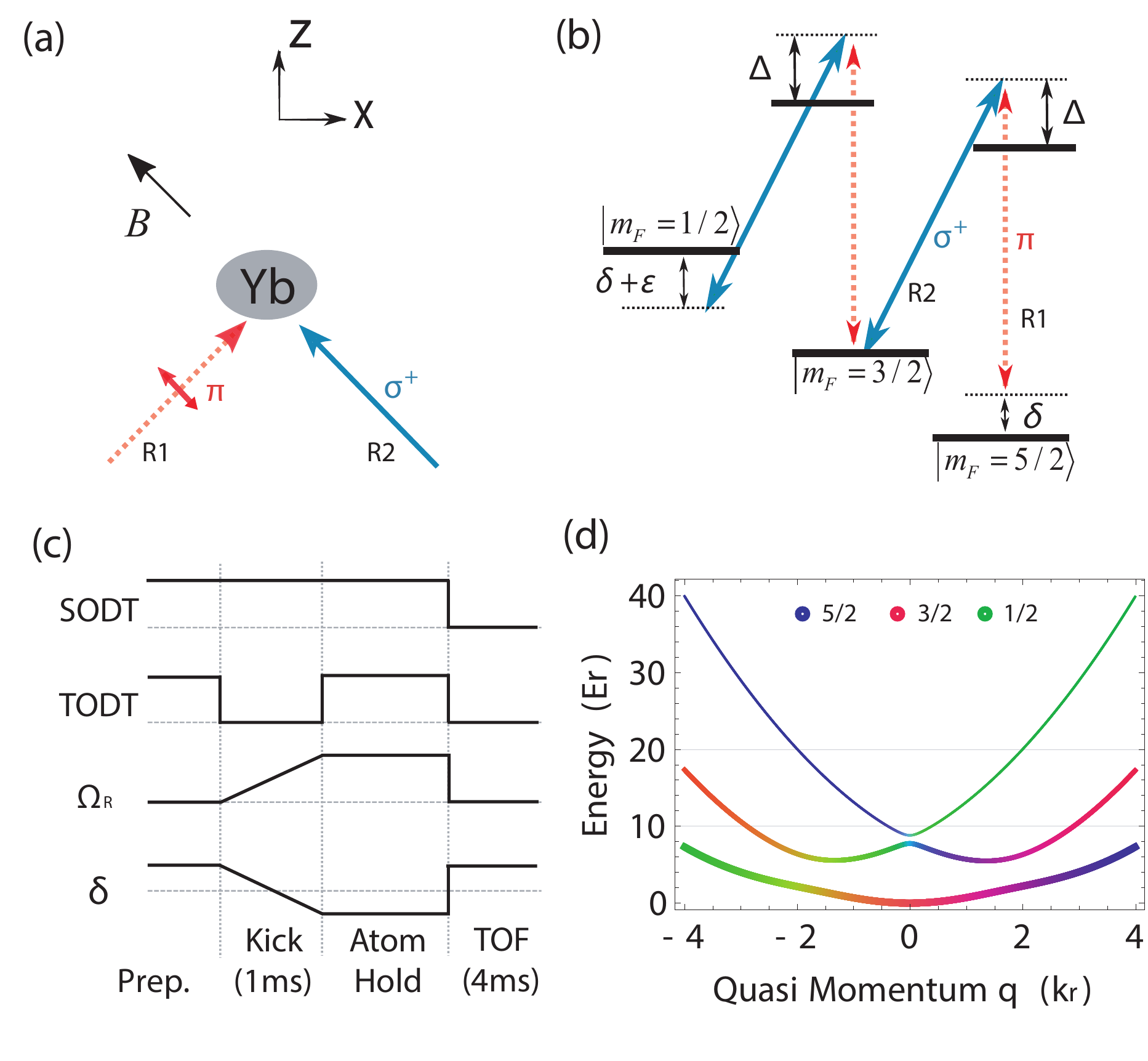}
\caption{\label{fig:1} \textbf{Experimental setup for realizing spin-1 SO coupling}. 
(a) Fermi gas of $^{173}$Yb atoms is initially prepared with a bias magnetic field along the $(-\hat{x}+\hat{z})$ direction and exposed to a pair of 556~nm beams ($R1$:linearly and $R2$:circularly polarzied respectly) intersecting the $\hat{z}$ axis at 45$^{\circ}$. 
(b) Two beams $R1$ and $R2$ consecutively couple the three ground-states hyperfine states $\ket{m_F=\frac{5}{2},\frac{3}{2},\frac{1}{2}}$ in the successive $\Lambda$ configuration. 
(c) The typical time sequence for inducing a dipole oscillation. A atomic cloud acquires an initial momentum along the $R2$ laser direction ($\hat{x}-\hat{z}$) after a brief switch off.  All Raman beams are kept on. The position of the atomic cloud is recorded after a TOF. (d) The typical single-particle energy dispersion in the quasi-momentum space for $\Omega_R=$4~$E_r$, $\delta=-$2.7~$E_r$ and $\epsilon=$5.3~$E_r$. The color of the band indicates the spin component as displayed in the inset.}
\end{center}\end{figure}

We begin experiments with a degenerate Fermi gas of $^{173}$Yb atoms by loading cold atoms pre-cooled in the intercombination magneto-optical trap into a crossed optical dipole trap (ODT). The crossed ODT, generated by 1064~nm lights,  consists of a tight ODT (tODT) in the horizontal plane and a shallow ODT  (sODT), the latter of which is tilted by 15 degree from the horizontal plane ~\cite{Song:2016fz,Song:2016ep}. In the middle of the forced evaporative cooling, we optically pump more than 70$\%$ atoms into the $\ket{m_F=\frac{5}{2}}$ state of the $^1S_0$ ground state by applying a weak nearly resonant 556~nm light with $\sigma^+$ polarization. A finite fraction of $\ket{m_F=\frac{3}{2}}$ state atoms are left to enhance the evaporative cooling and would be exhausted at the cooling process. After the final stage of the optical evaporative cooling, we achieve an almost single-component degenerate Fermi gas of $N=1.0\times 10^4$ atoms in $\ket{m_F=\frac{5}{2}}$ at $T/T_F\simeq 0.8$ together with $N=1.0\times 10^3$ atoms in $\ket{m_F=\frac{3}{2}}$, where $T_F$ is the Fermi temperature of the trapped atom with the trapping frequency of $\overline{\omega}=(\omega_x\omega_y\omega_z)^{\frac{1}{3}}$=$2\pi\times$126~Hz, $\omega_z \sim 2\pi\times$210~Hz and  $\omega_x\sim2\pi\times$150~Hz.

To realize SO coupling among three hyperfine states of $^{173}$Yb atoms, an external optical AC Stark shift is applied to separate out an effective spin-1 subspace that is composed of $\ket{m_F=\frac{5}{2},\frac{3}{2},\frac{1}{2}}$ from other hyperfine levels of the ground manifold. During the experiment, the magnetic field of 10~G is applied along the ($-\hat{x}+\hat{z}$) direction. A pair of beams, blue-detuned by $\sim$1~GHz from the intercombination $\lambda_0=$556~nm transition ${}^1$S$_0 (F = \frac{5}{2}) \leftrightarrow {}^3$P$_1 (F' = \frac{7}{2})$,  intersect at the atomic cloud with an angle of $\theta=$90$^{\circ}$, which imparts the momentum onto atoms and changes the energy dispersion as described in Fig.~\ref{fig:1}. As a result, we independently couple $\ket{m_F=\frac{5}{2}}\leftrightarrow\ket{m_F=\frac{3}{2}}$ and $\ket{m_F=\frac{3}{2}}\leftrightarrow\ket{m_F=\frac{1}{2}}$.

\begin{figure}[ptb]
\begin{center}
\includegraphics[width=8.9cm]{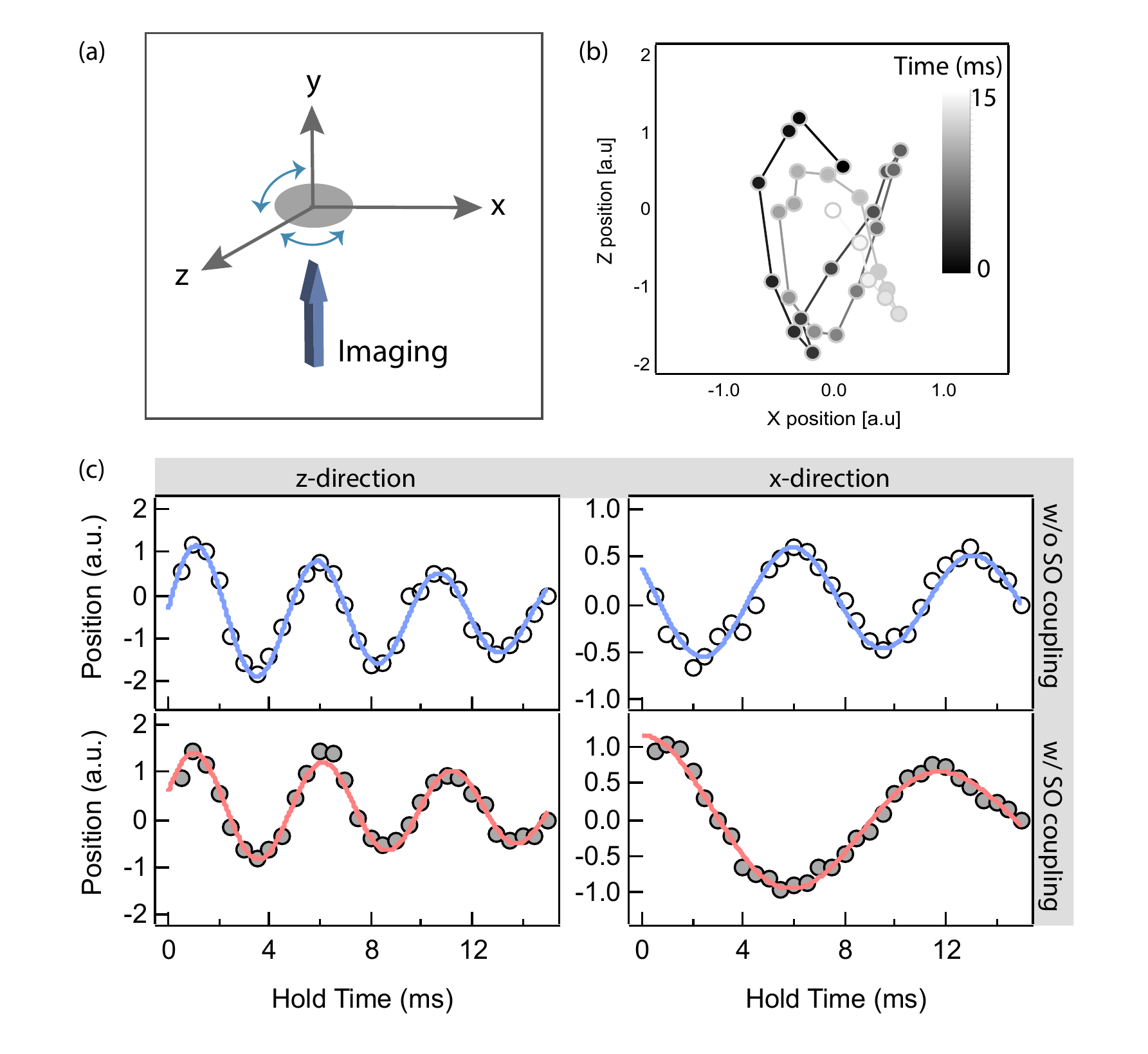}
\caption{\label{fig:2} {\bf Collective dipole mode with and without SO coupling}. (a) Induced dipole oscillations are monitored in both $\hat{x}$ and $\hat{z}$ directions after 4~ms TOF expansion. The absorption image is taken with a $^1S_0$-$^1P_1$ transition at 399~nm along the $\hat{y}$ direction. (b) A typical elliptical motion of the atomic cloud in the$x$-$z$ plane up to 15~ms after the initial kick. The color scale indicates the hold time after the collective mode is induced. (c) The dipole oscillation is not affected along the $\hat{z}$ direction that is perpendicular to the SO coupling direction. However, SO coupling significantly reduces the oscillation frequency along the $\hat{x}$ direction.}
\end{center}
\end{figure}

After the evaporative cooling in the crossed ODT, the two Raman coupling beams are linearly ramped up from zero intensity to the final value within 1~ms. The two-photon detuning $\delta$ is simultaneously ramped from $~\sim$50$E_r$ to the final value where $E_r$ is the recoil energy. We induce the collective dipole mode by briefly switching off the tODT beam for 1~ms, during which the atomic cloud is accelerated down by the gravity as described in Fig.~\ref{fig:1}(c). Subsequently the tODT is switch back on while the  Raman beam being kept on, followed by the variable hold time. With a weak probe light along the $\hat{y}$ direction, a time-of-flight (TOF) image is obtained by abruptly switching off all the ODT and Raman lights at the end of the hold time. After a 4~ms TOF ballistic expansion, we  record the momentum distribution of the atomic cloud and monitor the center of mass oscillation.

Raman transitions successively couple three hyperfine levels $\ket{F=\frac{5}{2},m_F=\frac{1}{2}}$, $\ket{F=\frac{5}{2},m_F=\frac{3}{2}}$ and $\ket{F=\frac{5}{2},m_F=\frac{5}{2}}$ in the $^1$S$_0$ ground manifold as described in Fig.~\ref{fig:1}(b). Each Raman beam induces the spin-dependent ac Stark shift. With an additional light lifting the degenacy, we separate out $\ket{F=\frac{5}{2},m_F=\frac{1}{2},\frac{3}{2},\frac{5}{2}}$ states from the remaining hyperfine levels, independently controlling both the detuning $\delta$ from the Raman resonance and the quadratic Zeeman shift $\epsilon$. We thus achieve the effective Hamiltonian in the quasi-momentum basis $\ket{m_F=\frac{5}{2}, k_x-2k_r}$, $\ket{m_F=\frac{3}{2}, k_x}$ and $\ket{m_F=\frac{1}{2}, k_x+2k_r}$ that describes spin-1 SO coupling:

%$$ H_{SOC}=\frac{1}{2m}(p_x + k_r \sigma_z)^2 + \frac{\Omega_R}{2}\sigma_x+\frac{\delta}{2}\sigma_z$$
\begin{equation*}
H_{SOC}=\hbar
\begin{pmatrix}
\frac{(q_x - 2 k_r)^2}{2m}-\delta & \Omega_R/2 & 0 \\
\Omega_R^*/2 & \frac{q_x ^2}{2m}& \tilde{\Omega_R}/2 \\
0 & \tilde{\Omega_R}^*/2 &  \frac{(q_x + 2 k_r)^2}{2m}+(\delta+\epsilon)
\end{pmatrix}
\end{equation*}

 where $q_x$ is the quasi-momentum of atoms along $\hat{x}$-direction, $k_r$ is the recoil momentum $k_r=k_0 \sin(\theta/2)$ for $k_0=2\pi/\lambda_0$, $m$ is the mass of ytterbium atom, $\Omega_R$($\tilde{\Omega}_R$) is the Rabi frequency of the Raman coupling for $\ket{m_F=\frac{5}{2}}\leftrightarrow\ket{m_F=\frac{3}{2}}$ ($\ket{m_F=\frac{3}{2}}\leftrightarrow\ket{m_F=\frac{1}{2}}$). With a given single-photon detuning of $\sim$1~GHz, we can approximate $\Omega_R=\tilde{\Omega}_R$ with current experimental setting. All values of detuning can be tuned by using spin-dependent light shifts induced by Raman beams and far-detuned beams that co-propagate with the beam $R2$ (see Fig.~\ref{fig:1}(a)).  Spin-dependent level shifts from different lights are  independently calibrated by monitoring the two-photon Rabi oscillation between $\ket{F=\frac{5}{2},m_F=\frac{3}{2}}$ and $\ket{F=\frac{5}{2},m_F=\frac{5}{2}}$, induced by a pair of co-propagating Raman beams~\cite{Song:2016ep}.

%by which the momentum of atoms was changed by 2$\hbar k_r$ imparting each Raman process with the recoil momentum is $k_r=k_0 \sin(\theta/2)$ for $k_0=2\pi/\lambda_0$. 
%where $p_x$ is the quasi-momentum of atoms along $\hat{x}$-direction, $m$ is the mass of ytterbium atom, $\Omega_R$ is the Rabi frequency of Raman coupling between 
%$\ket{F=\frac{5}{2},m_F=\frac{3}{2}}$ and $\ket{F=\frac{5}{2},m_F=\frac{5}{2}}$, $\delta$ is the two-photon detuning and $\epsilon$ is the differential AC stark shift.  

%\bigskip\noindent\textbf{\textsf{Implementation of SOC in Ytterbium173 - light shift for each $m_F$.}}\\
%{\color{red} In Fig.1, please replace (c) by the $m_F$-dependent light shift graph.}

Let us note that a relatively long-lived spin-orbit-coupled Fermi gas is a critical ingredient to monitor a collective dipole mode in the strong coupling regime around $\Omega_R\sim$4$E_r$. A narrow optical transition, readily available in ytterbium atoms~\cite{Song:2016ep} or other lanthanides~\cite{Burdick:2016jt}, allows to create SO coupling with minimal light-induced heating. In contrast to fermionic alkali atoms, alkaline-earth-like atoms have effective large fine-structure splitting for the triplet  $^3$P$_1$ excited state with a narrow line width (e.g.the natural linewidth of 182~kHz for $^{173}$Yb atoms). In our experimental setting, the 1~GHz detuning of the Raman beam still limits the lifetime upto $\sim$50~m in the strong coupling regime but much longer lifetime is expected with larger detuning in the red-detuned regime.

\begin{figure}%[ptb]
\begin{center}
\includegraphics[width=8.7cm]{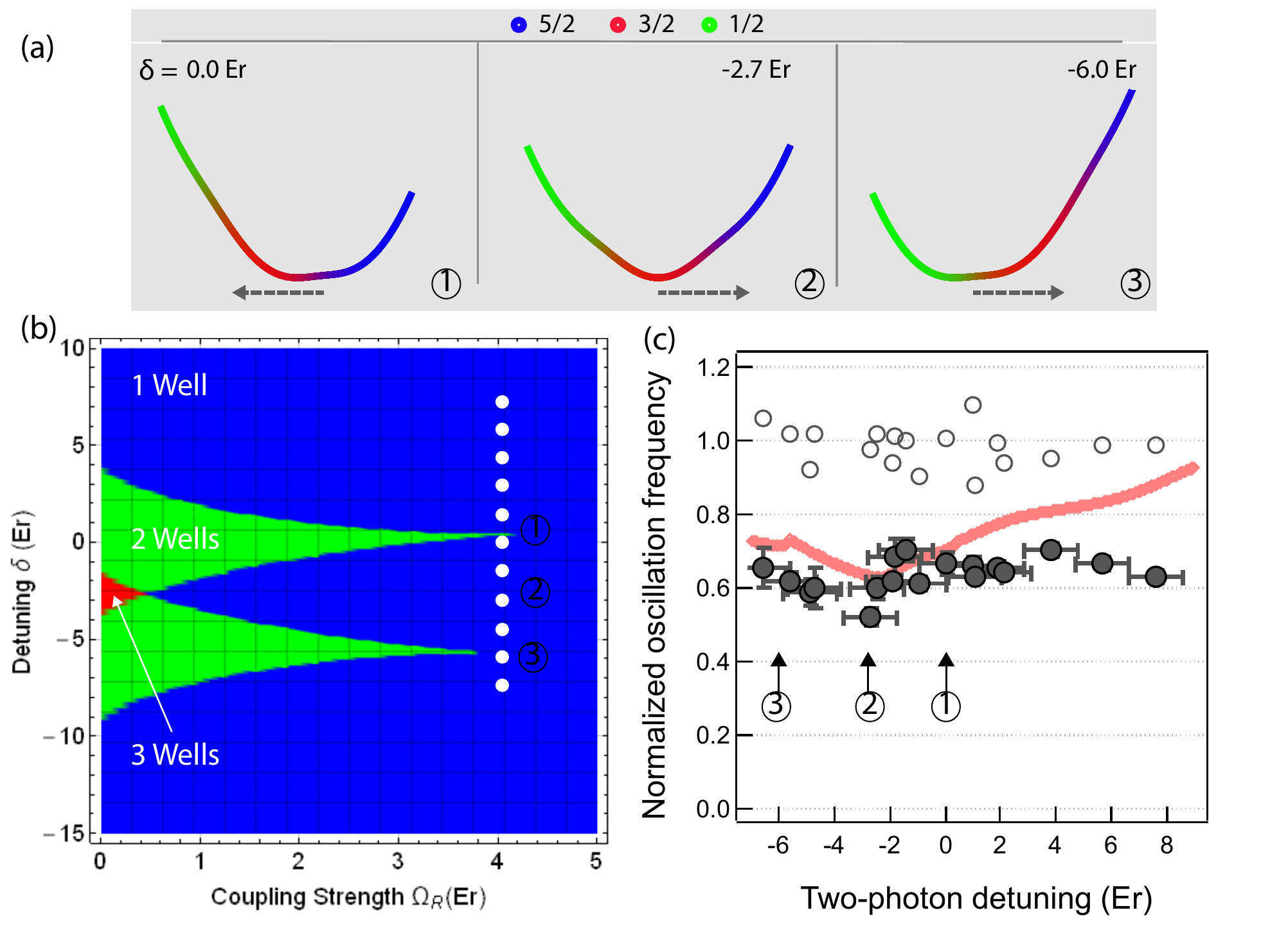}
\caption{\label{fig:3} \textbf{Measurement of the dipole oscillation frequency of spin-1 SO coupled fermions in the strong coupling regime}. The schematic phase diagram in the $\delta-\Omega_R$ plane is shown in (b). In the strong coupling regime at $\Omega_R=$4~$E_r$, multiple band minima, corresponding to three spin states, merge into the single minimum when the consecutive Raman coupling is identical around $\delta\sim-$2.7~$E_r$ as described in $\textcircled{2}$. When only two consecutive spins, $\ket{\frac{5}{2}}$ and $\ket{\frac{3}{2}}$, are resonantly coupled, the third spin $\ket{\frac{1}{2}}$ does not play the role at the bottom of the dispersion (see $\textcircled{1}$ in (a)). In (a), the dashed arrow indicates the initial momentum kick of $\sim$3$k_r$ applied to atoms. The dipole oscillation frequency is monitored along the $\hat{x}$ and $\hat{z}$ axes, the former of which has SO coupling. The oscillation is not affected along the $\hat{z}$ direction (open circle) while the oscillation becomes slower in the presence of SO coupling (solid circle). For this measurement, the Raman coupling is set to $\Omega_R=$4~$E_r$. The solid curve is the numerical prediction based on the semi-classical calculation.}
\end{center}
\end{figure}

%\bigskip\noindent \textbf{Measurement of the dipole oscillation frequency}\\
The collective dipole mode is a simple center-of-mass motion which usually exhibits the mode frequency same as the harmonic trap frequency according to the Kohn theorem~\cite{Kohn:1961et}. With SO coupling that breaks the Galilean invariance, however, a wave function of the particle  becomes sensitive to the velocity of particles through spin-momentum locking, which affects the frequency of the dipole mode. In the experiment, we start with a degenerate Fermi gas prepared at rest in the crossed  ODT.  The perturbation induced by the tODT generates a irregular elliptical motion in the $\hat{x}-\hat{z}$ plane after a TOF expansion (see Fig.~\ref{fig:2}(b)). The oscillation along the $\hat{z}$ and $\hat{x}$ are obtained by projecting the oscillation trace onto the corresponding axes, which are also two of three eigenaxes of the crossed ODT. The eigenaxes of the system are further confirmed by the principle component analysis so that a single frequency fit works for each axis. The collective mode frequency is then obtained by fitting the oscillation with a single frequency sinusoidal function with a exponential decay. All parameters are left as free parameters in the fitting process. To determine the reduced oscillation frequency with SO coupling, we independently measure the trap frequency of the crossed ODT with Raman beams at large two-photon detuning $\delta>50Er$. In this regime, the dispersion relation is almost the same as the one without SO coupling, but the lift beam or the Raman beam may slightly affect the bare trap frequency. The collective dipole mode is significantly affected along the $\hat{x}$ direction with SO coupling while the dipole oscillation frequency does not change along the $\hat{z}$ direction. Fig.~\ref{fig:2}(c) shows a typical dipole oscillation with and without SO coupling measured in the ODT.

The aforementioned oscillation measured so far has shown the capability of probing the collective dipole mode in the strong coupling regime. Here we set the Raman coupling strength $\Omega_R=$4~$E_r$ and scan the two-photon detuning $\delta$ from -7~$E_r$ to 8~$E_r$. In the schematic phase diagram of the $\delta-\Omega_R$ plane (Fig.~\ref{fig:3}(b)), the white dashed line shows the $\delta-\Omega_R$ parameters scanned in the experiment. For each point in the $\delta-\Omega_R$ plane, we obtain the normalized oscillation frequency as $\omega_i^{SOC}/\omega_i$ where $\omega_i^{SOC}$ ($\omega_i$) is the measured frequency with (without) SO coupling for $i=\{x,z\}$. The central result of our work is to identify the single minima dispersion relation that involves three spin states. Intuitively, the spin configuration adiabatically follows the dipole motion as long as atomic gases stay in the lowest energy band. When only two consecutive spin states are resonantly coupled (e.g. $\delta\sim-$6~$E_r$ or 0~$E_r$), the bottom of the dispersion curve is dominated by a spin-$\frac{1}{2}$ system as shown in Fig.~\ref{fig:3}(a). Around $\delta\sim-$2.7~$E_r$, three band minima corresponding to each spin state $\ket{\frac{5}{2},\frac{3}{2},\frac{1}{2}}$ merge into the single minimum at the bottom of the dispersion. Therefore, a sufficiently large initial momentum kick of $\sim$3~$k_r$ (indicated by dashed arrow in Fig.~\ref{fig:3}(a)) couples spin configuration with momentum during the dipole mode at $\delta=-$2.7~$E_r$ while only two spin states are coupled for $\delta=-$6~$E_r$ or 0~$E_r$.

In the experiment, we monitor collective dipole oscillations along two perpendicular axes with and without SO coupling after imparting an initial momentum kick. The dipole oscillation along the $z$ axis is not affected by SO coupling, which allows us to monitor and compensate systematic fluctuations of the trap frequency. Along the $\hat{x}$ direction, however, we observe the reduction of the dipole oscillation due to the presence of SO coupling. Around the detuning $\delta\sim-$2.7~$E_r$ where three spin states are symmetrically coupled, we observe the smallest value of the dipole oscillation frequency which is in good agreement with the prediction as follows.

%we aim to probe the effect of SO coupling in the strong coupling regime with a single minimum at the bottom of the band dispersion. In this strong coupling regime, we scan XXXX.
 %For this, we introduce SO coupling only along the $x$ direction (SO coupling direction), the eigenaxis of the harmonic trap, with the minimal effect on the $z$ direction (perpendicular direction). Fig.~\ref{fig:2} (c) shows typical collective dipole oscillations along two perpendicular axes with and without SO coupling. 
 %The simultaneous measurements of the dipole oscillations along both eigenaxes allows us to minimize systematic uncertainties and precisely determine the frequency of the collective dipole mode in the presence of SO coupling. 

%Note :anharmonicities reduce the measured oscillation frequency.

%we monitor the collective motion of fermions along two perpendicular eigenaxes of the harmonic trap, one of which is aligned to the SO coupling direction $x$. 

%The dipole mode oscillation of the atomic cloud are measured for different the two photon detunings when the spin orbital coupling strength $\Omega_R$ are fixed as 4$Er$. 

%\bigskip\noindent \textbf{Nemerical calculation}\\
For a quantitative comparison of the dipole oscillation to the measurement, we use a variational wave-function calculation assuming that the fermions stay in the lowest energy during the oscillation~\cite{PerezGarcia:1996gw}. We note that the higher bands may be populated during the initial momentum transfer when the Raman coupling is detuned by  $|\delta|>$10~$E_r$ in our system. In the experiment, however, the detuning $\delta$ is restricted within $\pm$8~$E_r$. Furthermore, the collisional rate $\tau^{-1}$ of the current system satisfies $\tau^{-1}<\omega_{x,y,z}$ and therefore we ignore collisional effects including spin-dependent interactions. This allows us to consider a spin-orbit-coupled Fermi gas in a semi-classical way as follows.

Our theoretical simulation is conducted based on the single particle Hamiltonian $H_{SOC}$ for the spin-1 manifold. With the collisionless approximation of the atomic gas, the oscillation of atomic could in momentum space can be numerically calculated from the dispersion relation of the lowest band $E(k_x)$ with the following set of equations:
%\begin{equation*}
$\frac{d k_x}{d t}=-\omega^2_{x} x$,$\frac{d x}{d t}=\frac{\partial E(k_x)}{\partial k_x}$
%\end{equation*}
where $\omega_x$ is the trap frequency along the SO coupling direction. Thus, it is straigthforward to have the equation $\frac{d^2 k_x}{d t^2}=-\omega^2_x \frac{\partial E(k_x)}{\partial k_x}$. By setting the intial value of $k_x$ as the momentum kick transferred to the atomic cloud, we thus simulate the evolution of the atomic cloud in momentum space. The momentum and time steps to run the numerical simulation are set to around 0.005~$k_r$ and 0.5~ms. To extract the frequency of numerically calculated oscillation, we conduct the fitting of the numerical results with a sinusoidal function with proper amplitude and initial fitting parameters. In this way, we calculate the  reduced dipole oscillation frequency in the presence of SO coupling as for the initial momentum kick of 3$k_r$ (see solid curve in Fig.~\ref{fig:3}(c)).  In the experiment, an atomic cloud has a finite momentum distribution due to the Fermi sea formed at the finite temperature.This momentum distribution not only causes the fast decay of the collective dipole oscillation but also enhance anharmonic behaviour associated. We attribute the deviation of the observed frequency from the prediction to the the anharmonicity as observed in the previous work\cite{Zhang:2012fd}.

We investigate the collective dipole mode in a spin-orbit-coupled fermions for the first time. Using the Raman transition consecutively coupling three internal hyperfine states, the Raman-dressed spin-1 spin-orbit-coupled Fermi gas is experimentally realized. Owing to a narrow optical transition of $^{173}$Yb atoms, the collective dipole mode can be monitored in the strong coupling regime with a long-lived sample. The measured oscillation frequency of the dipole mode is compared with the semi-classical calculation around a single-minimum dispersion, at which the minimum oscillation frequency is observed as expected for the spin-1 case. Our observation of the dipole mode will pave the way towards the study of spin-orbit-coupled fermions in hydrodynamic regime~\cite{Lan:2014fka}. Furthermore, this work highlights the remarkable capabilities to study the effect of SO coupling with large spin $s>\frac{1}{2}$ in a degenerate Fermi gas.  The large spin system would allow us to explore unprecedented quantum phenomena including SU($N$) Fermi liquids~\cite{Yip:2013vr,Pagano:2014hy} and SU($N$) Hubbard model~\cite{Taie:2012tb,Hofrichter:2016iq}, which can be further enriched by spin-momentum locking.

In near-term experiments, the measurement of spin susceptibility is conceivable by monitoring the spin oscillation over the dipole oscillation~\cite{Li:2012ef}. In a lanthanide cold atom system where inter-particle interaction strength can be easily tuned~\cite{Burdick:2016jt}, our present work shall push forward the future studies of interacting spin-orbit-coupled states, which are broadly discussed in theory but very hard to investigate in solid-state materials.

%\bibliography{allref}

%%%%-----------------
\section*{Acknowledgements}
G.-B. J. acknowledges the generous support from the Hong Kong Research Grants Council and the Croucher Foundation through ECS26300014, GRF16300215, GRF16311516, and GRF16305317, GRF16304918, C6005-17G and the Croucher Innovation grants respectively. G.-B. J also acknowledges the support from SSTSP at HKUST.

\end{document}